\documentclass[10pt,conference]{IEEEtran}
\IEEEoverridecommandlockouts
\pdfoutput=1
\usepackage{cite}
\usepackage{amsmath,amssymb,amsfonts}
\usepackage{float}
\usepackage{graphicx}
\usepackage{textcomp}
\usepackage{xcolor}
\usepackage{amsmath}
\usepackage{}
\usepackage{overpic}
\usepackage{verbatim} 
\usepackage{pifont}
\usepackage{times,amsmath,color,amssymb,graphicx, epstopdf, epsfig,cite,psfrag,subfigure,algorithm}
\usepackage[noend]{algpseudocode}
\usepackage{enumerate}
\usepackage{xcolor}
\def\BibTeX{{\rm B\kern-.05em{\sc i\kern-.025em b}\kern-.08em
    T\kern-.1667em\lower.7ex\hbox{E}\kern-.125emX}}
\IEEEoverridecommandlockouts\IEEEpubid{\makebox[\columnwidth]{ 978-1-6654-3540-6/22/\$31.00~\copyright~2022 IEEE \hfill} \hspace{\columnsep}\makebox[\columnwidth]{ }}
\begin{document}

\title{
		Optimal Job Scheduling and Bandwidth Augmentation in Hybrid Data Center Networks
{\footnotesize 
}
\thanks{This work is supported by the Natural Science Foundation of China (61931017). The corresponding author is Zhou Zhang.}
}

\author{\IEEEauthorblockN{Binquan Guo$^*$$^\dagger$, Zhou Zhang$^\dagger$, Ye Yan$^\dagger$, Hongyan Li$^*$}\\
	\IEEEauthorblockA{$^*$State Key Laboratory of Integrated Service Networks, Xidian University, Xi'an P.R.China\\
		$^\dagger$Tianjin Artificial Intelligence Innovation Center (TAIIC), Tianjin, P. R. China\\
		Email: bqguo@stu.xidian.edu.cn, yanye1971@sohu.com, zt.sy1986@163.com, hyli@xidian.edu.cn}}

\maketitle

\begin{abstract}
Optimizing data transfers is critical for improving job performance in data-parallel frameworks. In the hybrid data center with both wired and wireless links, reconfigurable wireless links can provide additional bandwidth to speed up job execution. However, it requires the scheduler and transceivers to make joint decisions under coupled constraints. In this work, we identify that the joint job scheduling and bandwidth augmentation problem is a complex mixed integer nonlinear problem, which is not solvable by existing optimization methods. To address this bottleneck, we transform it into an equivalent problem based on the coupling of its heuristic bounds, the revised data transfer representation and non-linear constraints decoupling and reformulation, such that the optimal solution can be efficiently acquired by the Branch and Bound method. Based on the proposed method, the performance of job scheduling with and without bandwidth augmentation is studied. Experiments show that the performance gain depends on multiple factors, especially the data size. Compared with existing solutions, our method can averagely reduce the job completion time by up to $10\%$ under the setting of production scenario. 

\end{abstract}


\section{Introduction}
%
%
%
%

Data transfer has a significant impact on application performance in data-parallel computing frameworks such as MapReduce \cite{dean2008mapreduce}, Pregel \cite{malewicz2010pregel} and Spark \cite{zaharia2012resilient}. These computing frameworks all implement a data partitioning model, in which jobs are decomposed into finer-grained tasks, and massive amounts of intermediate data between their computation stages need to be transferred through the network before generating the final results.
For many applications in production environment, 
the data transfers account for more than 50\% of the job completion times \cite{chowdhury2011managing}. 
With the rapid growth of the processed data size, 
the network resource has become an increasingly significant bottleneck in the performance of cloud computing. 

Traditional data center networks (DCNs) which consist of copper and optical fiber cables provision the link capacity between racks in a fixed manner. During a job's execution, however, data flows trend to be bursty when multiple tasks are ready for data transmission and hence exhibit dynamic patterns. When the traffic between two racks exceeds the provisioned capacity, congestion will occur. Such static link capacity allocation restrict the support of parallel data transfer and therefore slow down the subsequent tasks' execution duration the job execution.

To support the dynamic allocation of network resources, many efforts have recently been made to deploy the wireless communication technologies into wired DCNs to enable dynamic bandwidth augmentation, such as mmWave links \cite{terzi202160} and free-space optics (FSO) \cite{celik2019optical}. 
60GHz antennas and FSO transceivers can provide Gigabit transmission capability with low-latency switching time.
By leveraging mmWave MIMO beamforming,  a large number of beams can be scheduled with extremely small switching delay \cite{abari2016millimeter}. And the reconfiguration delay of FSO was shown to be only 12 $\mu s$ while supporting 18,432 fanouts \cite{ghobadi2016projector}.
As a result, these reconfigurable wireless technologies demonstrate the potential for providing additional bandwidth by dynamically
establishing wireless links on demand to offload traffic and reduce the job completion time.

\textcolor{black}{In order to intuitively show both the advantages and challenges of using wireless transmission for reducing job completion time, an example job consists of five tasks is presented in Fig. 1. Assume the transmission capacity of all wired links and wireless transceivers between racks are 10 Gbps. 
	With only wired links, the intermediate data during each stage must be transmitted sequentially, resulting the prolonged job completion time. 
	By using dynamically established wireless links to transmit data on \textit{task$_1$} $\longrightarrow$ \textit{task$_4$} and \textit{task$_2$} $\longrightarrow$ \textit{task$_5$}, $16\%$ of the job completion time can be reduced.}
Thus, an appropriate wireless bandwidth augmentation scheme can greatly speed up the job execution.
However, it also requires the job scheduler and transceivers to make joint decisions under coupled computing and communication constraints.
		\begin{figure}
	\centering
	\includegraphics[width=81mm]{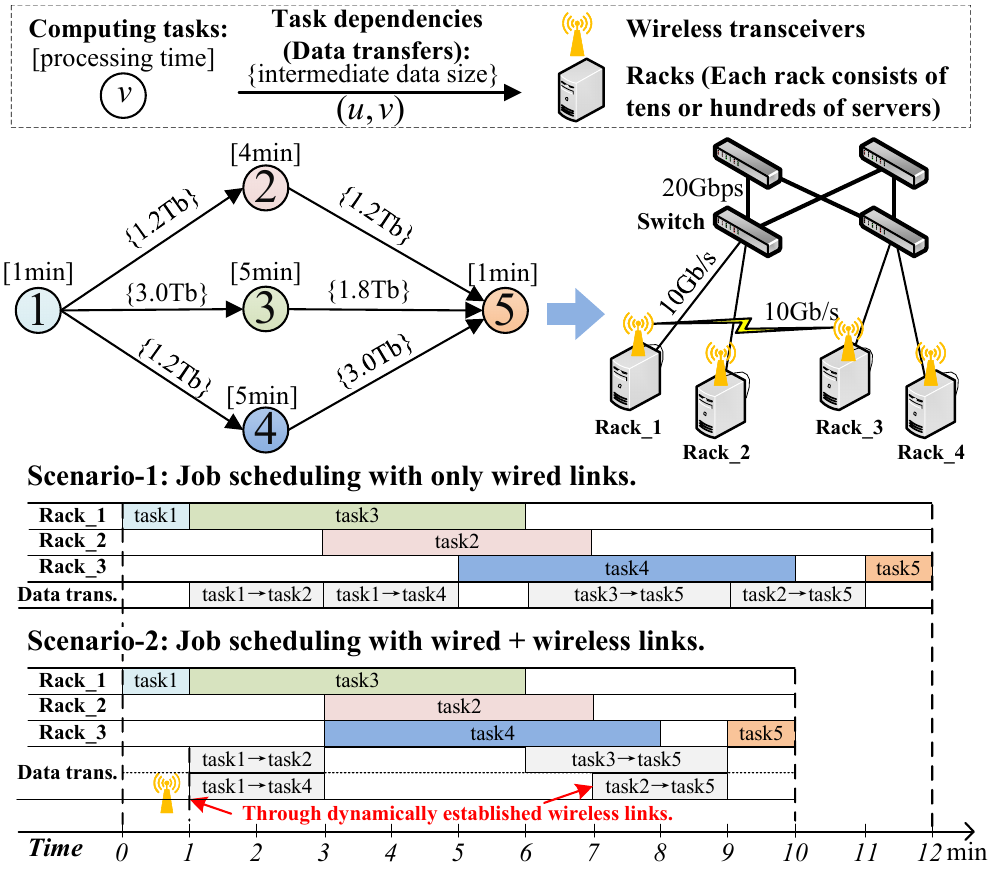}
	\caption{An example to illustrate the advantages and challenges of using dynamically established wireless links to reduce job completion time.}
\end{figure}

In our previous work \cite{luo2019energy}, a flow routing and antenna scheduling scheme is proposed for hybrid DCNs without considering computing tasks. There are many important works focus on enhancing flow scheduling performance using wireless technologies, with the aim of minimizing the network congestion \cite{han2015rush}, relieving hotspots \cite{halperin2011augmenting}, enhancing the network flow throughput \cite{cui2013dynamic}, or reduce the length of flow paths \cite{li2019energy}.
However, these studies assume that the computing tasks have already been assigned and hence the endpoints of flows are predetermined,
without jointly scheduling the computation and communication.
The most related work to ours is \cite{ao2021joint}. 
Reference \cite{ao2021joint} studies the joint wireless links scheduling and computing task assignment problem and obtains substantial performance gain. However, its model assumes tasks are independent and can be processed simultaneously, without considering dependency constraints between adjacent tasks.

In this work, we aim to jointly schedule dependency constrained tasks and wireless transceivers in hybrid DCN. 
We identify that such a problem is a complex mixed integer non-linear programming problem, which is not solvable by existing optimization methods. To overcome this, we transform it into an equivalent problem based on the coupling of its bounds, the revised data transfer model and non-linear constraints reformulation, such that the optimal solution can be acquired efficiently by the Branch and Bound method. 
Through numerical experiments, we find the performance gain introduced by wireless augmentation depends on multiple factors, especially the data size.
Compared with existing solutions, our method can averagely reduce the job completion time by up to $10\%$ under the setting of production scenario.




\section{System Model}

Consider a hybrid DCN consists of a set of racks.
Each rack is composed of a number of servers for computation and storage, and equipped with reconfigurable wireless transceiver for bandwidth augmentation. 
The racks are connected with both the wired links with fixed capacity and the dynamically established wireless links. 
We assume the orthogonal channel allocation and progressive directional antenna are used, such that the total wireless bandwidth is shared by wireless links among racks via FDMA without interference.

In this work, we consider periodic jobs, which are loaded everyday and their detailed knowledge can be profiled from historical logs\footnote{According to \cite{ren2012workload}, periodic jobs can be optimized and account for $80\%$ of the workload in Hadoop cluster at Taobao.}. Each job is described by a directed acyclic graph (DAG) ${G}=(\mathcal{V},\mathcal{E})$, as in job scheduling systems like Fuxi \cite{zhang2014fuxi}. $\mathcal{V}$ is the set of computing tasks, and $\mathcal{E}$ is the set of directed edges representing the dependency between adjacent tasks. 
Each task $v \in \mathcal{V}$ specifies its unit size of resources, e.g., \{1 core CPU, 1GB Memory\}, thus its processing time can be measured as $p_v$. Each edge $(u, v) \in \mathcal{E}$ specifies the data size $d_{(u, v)}$ from task $u$ to task $v$. 
The required bandwidth of transmitting data across racks is specified as $B_s$.

Upon receiving the job, the job scheduling system will check the free resources among racks, and try to allocate computing and bandwidth resources which meet the job's resource requirements.
Let $ \mathcal{M}=\{ 1,2,...,M \}$ be the set of feasible racks. 
\textcolor{black}{The allocated wired bandwidth between each pair of racks must be guaranteed as $B_s$.} 
For wireless resources, the available wireless bandwidth is divided into multiple orthogonal subchannels denoted by a set $\mathcal{K}$, and each subchannel ${k} \in \mathcal{K}$ has a bandwidth of $B$.
With the allocated bandwidth resources, 
the transferring time of the data on edge $(u,v) \in \mathcal{E}$ through wired links is calculated as $q_{(u,v)}= \frac{d_{(u,v)}}{B_{s}}$, and the transferring time through wired links is calculated as $\check{q}_{(u,v)}=\frac{d_{(u,v)}}{B}$.   
Otherwise, if adjacent task $u$ and $v$ are assigned to the same rack, the delay of transferring the data locally is denoted as $r_{(u,v)}$.

\section{Problem Formulation} 
\subsection{Common Constraints for Computing Task Assignment} 
We define the binary variable ${x}_{vi}$ and the continuous variable $s_v$ for each task $v \in \mathcal{V}$.
Specifically, ${{x}_{vi} = 1 }$ means task $ {v}$ is assigned to rack ${i}$, and $s_v$ denotes task $v$'s start time.
Inherently, the following constraints must be satisfied:

\subsubsection{Non-repetition Constraints}Each task $v \in \mathcal{V}$ must be assigned to one rack and processed only once,
namely,  
\begin{equation}
\sum_{i \in \mathcal{M}} {x}_{vi} = 1, \forall v \in {\mathcal{V}}.
\end{equation}
\subsubsection{Non-preemption Constraints}To prevent computing resource overload, each rack is allowed to process the job's one task at a time. Once started, a task cannot be interrupted by any others until its completion. For $\forall v,v' \in {\mathcal{V}}, v \neq v'$, 
\begin{equation}\label{computing_non_overlap}
s_v +  p_{v} \leq  s_{v'} \text{ \emph{or} }  s_{v'} +  p_{v'} \leq  s_{v},\text{\emph{if}}  \underbrace{\sum_{i \in \mathcal{M}} i {x}_{vi} = \sum_{i \in \mathcal{M}} i {x}_{v'i}}_{\mathbf{C1}}.
\end{equation}
Expression  $ \sum_{i \in \mathcal{M}} i {x}_{vi} $, $ \sum_{i \in \mathcal{M}} i {x}_{v'i} $ in \textbf{C1} represent the selected rack for task $v$, $v'$, respectively. Constraint (\ref{computing_non_overlap}) guarantees if two computing tasks $v$ and $v'$ are assigned to the same rack, there is no resource competition between them.

\subsubsection{Precedence Constraints} A computing task only starts after the completion of all its precedent tasks, namely, 
\begin{equation}\label{computing_precedence}
s_u  + p_u \leq  s_v, \forall (u,v) \in \mathcal{E}.
\end{equation}
\textit{Remark 1:} Note that constraint (\ref{computing_precedence}) is relatively slack, due to the fact that it ignores the data transfer time between adjacent tasks, which will be discussed in the next subsection.

\subsection{Constraints for Intermediate Data Transfers}

Coupled with the assignment decisions of computing tasks, the intermediate data among tasks may be transfered locally without occupying cross-rack links, or be transmitted externally through either wired or wireless links. 
For clarity, we define binary variable ${z}_{(u,v)}$ for each edge $(u,v)$, namely,  
\begin{equation}
{z}_{(u,v)}:=0, \text{ \emph{if} }  \sum_{i \in \mathcal{M}} i {x}_{ui} = \sum_{i \in \mathcal{M}} i {x}_{vi},\forall (u,v) \in \mathcal{E},
\end{equation}
where ${z}_{(u,v)}=0$ means task $u$ and $v$ are assigned to the same rack.
In this case, the data on edge $(u,v)$ will be transferred locally (i.e., within a rack) with delay $r_{(u,v)}$, namely, 
\begin{equation}
s_u  + p_u + r_{(u,v)} \leq  s_v, \forall (u,v) \in \mathcal{E}, \text{ \emph{if} } {z}_{(u,v)} =0.
\end{equation}
Otherwise, if ${z}_{(u,v)}=1$, task $u$ and $v$ will be assigned to different racks, and the \textit{network flow between racks} will occur.

\textbf{Heterogeneous network flow scheduling constraints}: 
We define binary variables $\alpha_{(u,v)}$ and ${y}_{(u,v),k}$, 
 in which $\alpha_{(u,v)}=1$ means the data on edge $(u,v)$ is transferred via wired links. ${{y}_{(u,v),k} = 1}$ means the data is assigned to wireless subchannel ${k}$. 
The start time of data transmission from task $u$ to $v$ is denoted as the continuous variable $s_{(u,v)}$.
Firstly, the data on edge $(u,v)$ can only start to be transmitted until the completion of computing task $u$, namely, 
\begin{equation}
s_u  + p_u \leq  s_{(u,v)}, \forall (u,v) \in \mathcal{E}, \text{ \emph{if} } {z}_{(u,v)}= 1.
\end{equation}

\subsubsection{Data Transmitted Through Wired Links} 
If the data on edge $(u,v)$ is transmitted through wired links, 
the subsequent task $v$ can only start after it receives all the data, namely, 
\begin{equation}
s_{(u,v)} + q_{(u,v)}  \leq  s_v, \forall (u,v) \in \mathcal{E}, \text{ \emph{if} } {z}_{(u,v)} =\alpha_{(u,v)} = 1.
\end{equation}
To prevent congestion, for each pair of different network flow $(u,v)$ and $(u',v')$ transferred via wired links, there is
\begin{equation}
s_{(u,v)} + q_{(u,v)}  \leq  s_{(u',v')} \text{ or } s_{(u',v')} + q_{(u',v')}  \leq  s_{(u,v)}, 
\end{equation}
where ${z}_{(u,v)} =\alpha_{(u,v)} = {z}_{(u',v')} =\alpha_{(u',v')}= 1$ is required.

\subsubsection{Data Transmitted Through Wireless Links}  Similarly, if the data is transmitted through wireless subchannels, the subsequent task must wait until the data transfer ends. $\forall (u,v) \in {\mathcal{E}} $, 
\begin{equation}
s_{(u,v)} + \check{q}_{(u,v)}  \leq  s_v, \text{ \emph{if} }  {z}_{(u,v)} =1 \text{ \emph{and} } \alpha_{(u,v)} = 0.
\end{equation}

To prevent wireless interference, each subchannel is allowed to transfer one network flow during any period of time, and once started,
the data transmission cannot be interrupted until its completion. For $\forall (u,v), (u',v') \in {\mathcal{E}}, (u,v) \neq (u',v')$,
\begin{equation}\label{communication_non_interference}
\begin{split}
& s_{(u,v)} +  q_{(u,v)} \leq  s_{(u',v')} \text{ or }  s_{(u',v')} +  q_{(u',v')} \leq  s_{(u,v)}, 
\\ & \text{ \emph{if} }  \underbrace{\sum_{k \in \mathcal{K}} k {y}_{(u,v),k}  = \sum_{k \in \mathcal{K}} k {y}_{(u',v'),k}}_{\mathbf{C2}}
\text{ \emph{and} }   {\alpha}_{(u,v)} = 0.
\end{split} 
\end{equation}
Expression  $ \sum_{k \in \mathcal{K}} k {y}_{(u,v),k} $, $ \sum_{k \in \mathcal{K}} k {y}_{(u',v'),k} $ in \textbf{C2} indicate the selected subchannel for transferring data on edge $(u,v)$, $(u',v')$, respectively. Therefore, constraint (\ref{communication_non_interference}) guarantees if two network flows are transferred over the same subchannel, there is no interference during their transmission.

\subsection{Problem Formulation}
The objective is to minimize the job completion time. Thus the original problem can be formulated as follow, 
\begin{equation*}
\begin{split}
\mathbf{OP: }\mathop  \text{ min }\limits_{\mathbf{s},\mathbf{x},\mathbf{y}}  &  \max \{ {s_v + {p_v}} \  | \  \forall v \in \mathcal{V}  \} \\ \text{s.t. } &  (1)-(10),
\end{split}
\end{equation*}
where $\mathbf{s} = \{ s_v, \forall v \in \mathcal{V}\} \cup \{s_{(u,v)}, \forall (u,v) \in \mathcal{E}\}$, $ \mathbf{x} = \{ x_{vi}, \forall v \in \mathcal{V}, \forall i \in \mathcal{M}  \} $ and $ \mathbf{y} = \{ y_{(u,v),k}, \forall (u,v) \in  \mathcal{E}, \forall k \in \mathcal{K} \} $.

It is observed that \textbf{OP} is a complex Mixed Integer Non-linear Programming (MINLP) with a large number of coupled constraints,  which is not directly solvable by existing optimization methods. 
The exhaustive search for the optimal solution is intractable, due to the huge solution space imposed by logical and disjunctive constraints.
Even for a common scale \textbf{OP} (e.g., \textbf{job size $\leq 10$} in production cases\cite{ren2012workload}), searching for the optimal solution is non-trivial, and the time complexity is unacceptable. In the next section, we will transform \textbf{OP} into an  equivalent problem based on combination of multiple steps, which paves the way for adopting the sophisticated optimization methods to acquire its optimal solution efficiently.

\section{The Optimal Job Scheduling and Bandwidth Augmentation Scheme}

To make it possible to solve \textbf{OP} within a reasonable time, we first use heuristics to estimate its upper and lower bound. Next, we introduce the generalized data transfer model to linearize the coupled constraints between task assignment decisions and data transfers. Then, disjunctive reformulation technique combined with multiple auxiliary variables is adopted to convert the resource constraints of problem \textbf{OP} into their linearized forms, which allows us to acquire its optimal using the Branch and Bound method.

\subsection{\textcolor{black}{Heuristic-based Bounds Estimation}}
\textbf{Upper Bound: }For any given job, a feasible scheduling scheme can be obtained by assigning all its tasks to a single rack. In this case, tasks are processed in a topological sort order without cross-rack data transmission. Thus its job completion time can be calculated as $T_{\emph{max}} = \sum_{v \in \mathcal{V}} p_v + \sum_{(u,v) \in \mathcal{E}} r_{(u,v)}$. We define $T_{\emph{max}}$ as the upper bound of \textbf{OP} by assuming any "good" scheduling schemes cannot be worser than this scheme, namely, $ \max \{ {s_v + {p_v}}  | \  \forall v \in \mathcal{V} \} \leq T_{\emph{max}} $.

\textbf{Lower Bound: } The lower bound of \textbf{OP} can be obtained by summing up the processing time of computing tasks and the local data transfer delays along the \textit{longest branch} of the given job. For simplicity of illustration, we present an example in Fig. 2. Fig. 2(a) is an example DAG job graph, while Fig. 2(b) is the converted cost graph by transforming all of the node's cost into their outgoing edge's costs. 
Then the longest path algorithm can be used to calculate the distance from start node to each task $v$ (i.e., the earliest start time of task $v$) as $dist(v)$. Finally, the longest branch length can be obtained as $ T_{\emph{min}} = \max_{{v} \in \mathcal{V}} \{ dist(v) + p_v \}$, which is the lower bound of \textbf{OP}. The detailed procedure is presented in Algorithm 1. 

	\begin{figure}
	\centering
	\includegraphics[width=81mm]{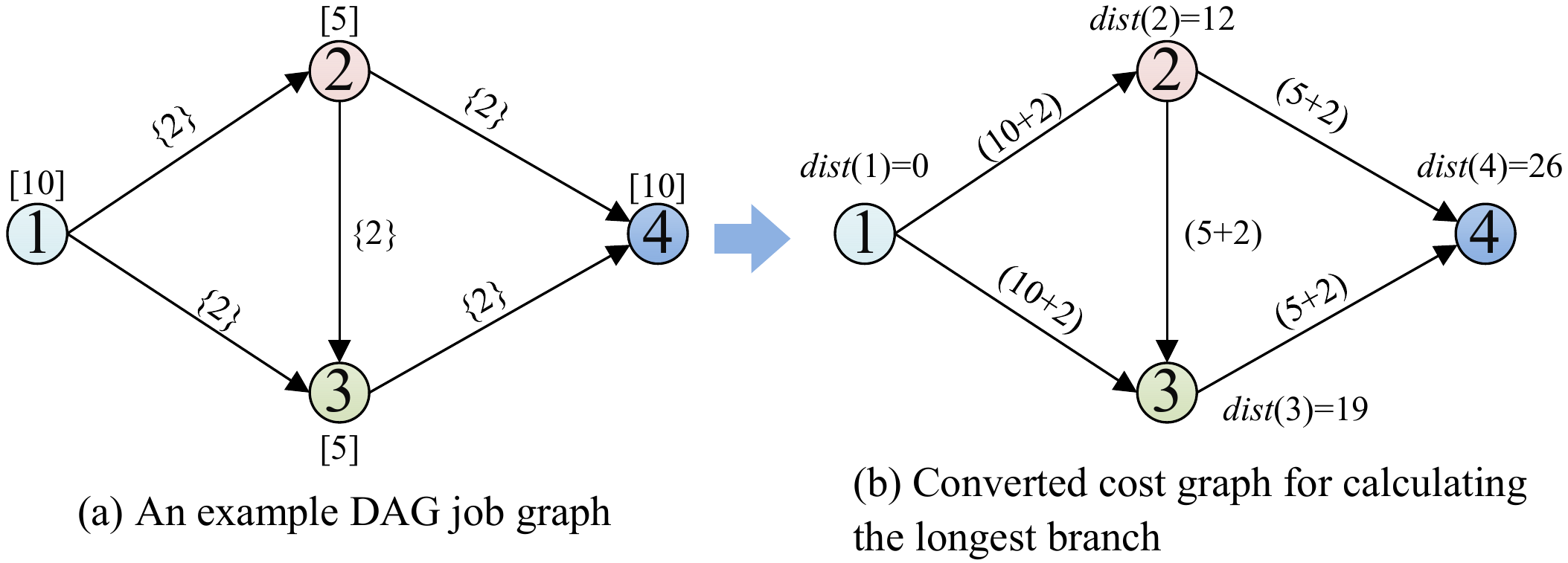} \
	\caption{An example for calculating the longest branch of DAG job graph.}
	\label{fig2}
\end{figure}
\begin{algorithm}
	\caption{The Longest Branch Algorithm}
	\label{array-sum}
	\hspace*{0.02in} {\bf Input:}
	\text{Job} $G=(\mathcal{V},\mathcal{E}), p_v, \forall v \in \mathcal{V}$, and $r_{(u,v)}, \forall (u,v) \in \mathcal{E}$.\\
	\hspace*{0.02in} {\bf Output:} The longest branch length of job $G$.
	\begin{algorithmic}[1]
		\item  Define $c_{(u, v)}$ as the cost of edge $(u, v) \in \mathcal{E}$.
		\item  \textbf{for} each task $v \in \mathcal{V}$ \textbf{do}
		\item  \ \ \ \  Initialize  $dist(v)=0$ as the distance from start to $v$.
		\item  \ \ \ \  \textbf{for} each outgoing edge $(v,x)$ of task $v$ \textbf{do}
		\item  \ \ \ \ \ \ \ \   Set  $c_{(v, x)} = p_v + r_{(v, x)}$.
		\item  Topologically sort $\mathcal{V}$ in $G$.
		\item  \textbf{for} each task $v \in \mathcal{V}$ in topological sort order \textbf{do}
		\item  \ \ \ \  Update $dist(v)= \max_{{(u, v)} \in \mathcal{E}} \{dist(v) + c_{(u, v)}\}$.
		\item  {\bf return} $\max_{{v} \in \mathcal{V}} \{ dist(v) + p_v \}$.
	\end{algorithmic}
\end{algorithm}

\subsection{\textcolor{black}{Generalized Representation of Data Transfer}}
		\begin{figure}
	\centering
	\includegraphics[width=81mm]{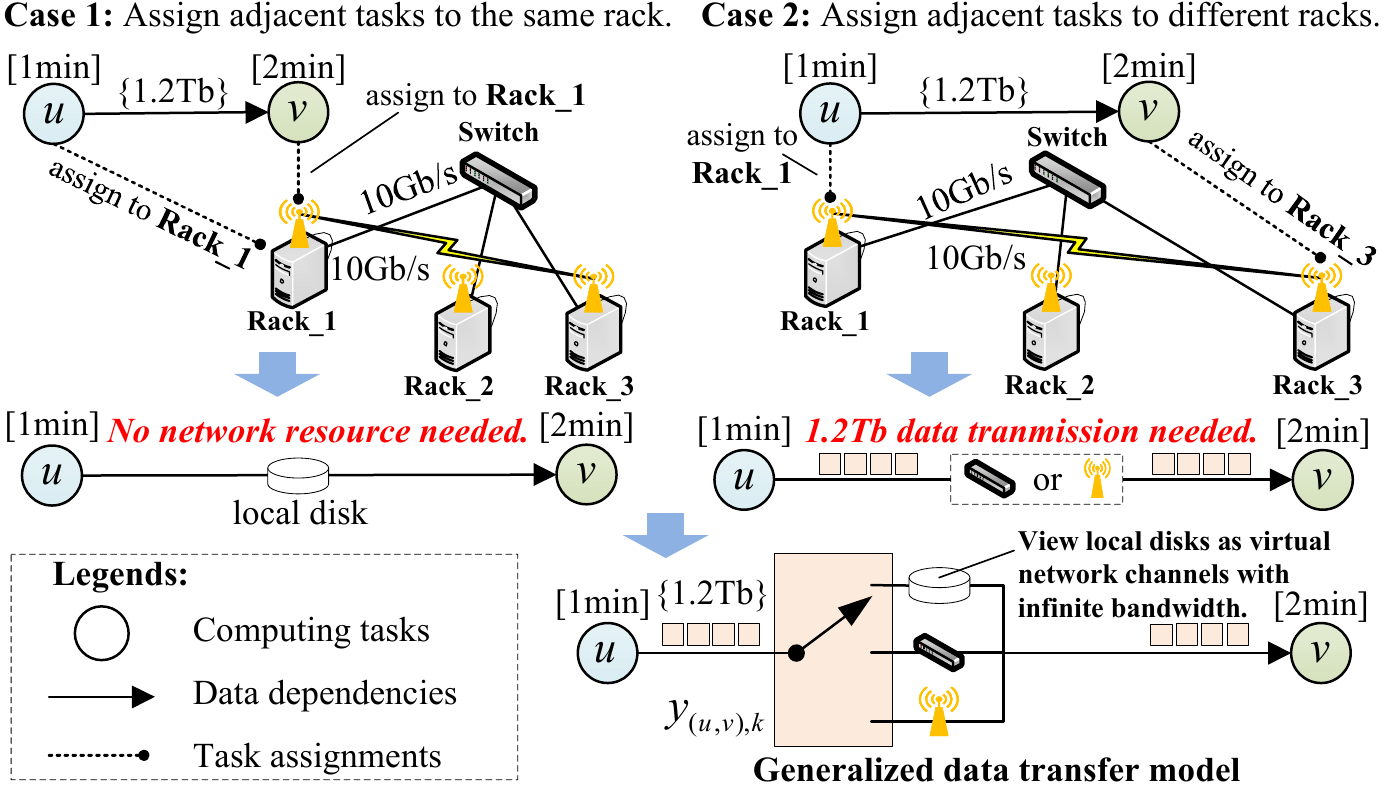} \
	\caption{Illustration of generalized data transfer model.}
	\label{fig2}
\end{figure}

Depending on the assignment decisions of adjacent tasks, the intermediate data on each edge between adjacent tasks is either available in local disks, or transferred through wired or wireless links.
To eliminate the logical constraints associated with variables $z_{(u,v)}$ and \textcolor{black}{cover different cases of data transfers}, we devise the generalized representation of data transfer model by introducing the virtual channel $c$ with infinite bandwidth and the wired channel $b$, in which each edge is associated with a "single pole triple throw switch".
As illustrated in Fig. 2, the case of local availability of data is viewed as data transmitted over an infinite channel without resource conflict but a constant delay, since there is no cross-rack data transmission needed when the adjacent tasks are assigned to the same rack.
Thus, for each intermediate data, it will be transferred through a channel from there types of network resources denoted by the set $\{b,c\}\cup \mathcal{K}$.
By adopting the generalized data transfer model, 
the intermediate data on edge $(u,v) \in \mathcal{E}$ must be transferred on one of communication channels from $\{b,c\}\cup \mathcal{K}$, namely,  
\begin{equation}\label{c_generalFlowCompletion}
\sum_{k\in \{b,c\} \cup \mathcal{K}} {{y}_{(u,v),k}} = 1,\forall (u,v) \in \mathcal{E},
\end{equation}
where ${{y}_{(u,v),k}}=1$ indicates the intermediate data on edge $(u,v)$ is transferred via communication channel $k$. 

Therefore, if the data on edge $(u,v)$ is transferred through wired links, $\sum_{k \in \{b\}} {y}_{(u,v),k} = 1$;
if the data is transferred through wireless subchannel, 
$ {y}_{(u,v),b} = 1$; 
otherwise, the data is transferred through local disk and then 
${y}_{(u,v),c} = 1$.

\subsection{Constraints Decoupling and Reformulation}
With the bounds and the generalized data transfer model, we can linearize OP based on the disjunction reformulation technique.
We define auxiliary variable $\tilde{x}_{vi}  \in [0,T_{\emph{max}}]$ for each ${{x}_{vi}}$, in which 
$ {\tilde{x}_{vi}  = \tau }$ denotes task $ {v}$ is assigned to rack ${i}$ and begins to process at time $ {\tau}$, \textcolor{black}{otherwise ${\tilde{x}_{vi}  = 0}$}. Similarly, we define $ \tilde{y}_{(u,v),k} \in [0,T_{max}] $ as auxiliary variable for
$ {y}_{(u,v),k} $  in which 
$ {\tilde{y}_{(u,v),k}  = \tau }$ denotes that the intermediate data on edge $ {(u,v)}$ is assigned to channel ${k} \in \{b,c\} \cup \mathcal{K}$ and begins to transfer at time $ {\tau }$, otherwise $ {\tilde{y}_{(u,v),k} = 0 }$. The following constraints can bind these variables, $\forall v \in \mathcal{V}, \forall e \in \mathcal{E}$,
\begin{equation}
\begin{split}
\tilde{x}_{vi} -1 \leq {{{x}}_{vi}} \cdot T_{\emph{max}} - (1-{{x}_{vi}}) \cdot \varepsilon, \forall i \in \mathcal{M}.
\end{split}
\end{equation}
\begin{equation}
\begin{split}
&  \tilde{y}_{ek} -1 \leq {{y}_{ek}}  \cdot T_{\emph{max}} - (1-{{y}_{ek}}) \cdot \varepsilon, \forall k \in \mathcal{K} \cup \{b,c\}.
\end{split}
\end{equation}
Here $T_\emph{max}$ acts as a big constant,
and $\varepsilon \in (0, 1)$ is a small constant commonly used in
the logical constraints reformulation of MINLP and can be set as 0.1 in practice.

Next, let  $ {\boldsymbol{\psi}_{vv'}}  \in \{0,1\}^{|\mathcal{M}|} $ be the task assignment indicator. Specifically, $ {\boldsymbol{\psi}_{vv'i}} = 1 $ indicates that task $v$ and $v'$ are assigned to the same rack $ i $, otherwise 
$  {\boldsymbol{\psi}_{vv'i}} = 0 $. Further, let ${\sigma}_{vv'} \in \{0,1\}$ be the precedence indicator such that, 
if task $v$ starts no later than $v'$, 
${\sigma}_{vv'} = 1$.
Similarly, for data transmission, we define the binary variables $\boldsymbol{\chi}_{ee'} \in \{0,1\}^{|\{b\}\cup {\mathcal{K}}|}$ as the contention indicator, $ {\boldsymbol{\chi}_{ee'k}} = 1 $ if the data on edge $e$ and $e'$ compete for the network channel $k$; and $\phi_{ee'}\in \{0,1\}$ as precedence indicator between network flows such that, if the data on $e$ begins to transfer no later than the data on $e'$, 
${\phi}_{ee'} = 1$, where $e, e' \in \mathcal{E}, e \neq e'$. Eventually, the following constraints are required to 
construct the indicator variables.
\begin{equation}
\sum_{i \in \mathcal{M}} \boldsymbol{\psi}_{vv'i} \leq 1, \forall v,v' \in \mathcal{V}, v \neq v'
\end{equation}
\begin{equation}
\sum_{k \in \mathcal{K}\cup \{b\}} \boldsymbol{\chi}_{ee'k} \leq 1, \forall e,e' \in \mathcal{E}, e \neq e'.
\end{equation}
\begin{equation}
\begin{split}
0 \leq {{x}_{vi}} + {{x}_{v'i}} - 2 \cdot \boldsymbol{\psi}_{vv'i} \leq 1, \forall i \in \mathcal{M}  
\end{split}
\end{equation}
\begin{equation}
\begin{split}
& 0 \leq {{y}_{ek}} + {{y}_{e'k}} - 2 \cdot \boldsymbol{\chi}_{ee'k} \leq 1,  \forall k\in \mathcal{K}\cup \{b\}
\end{split}
\end{equation}
\subsubsection{Computing resource constraint reformulation} 

To ensure the execution of any two computing tasks on the same rack does not overlap, the computing resource constraints can be 
linearized by utilizing disjunctive programming formulation technique
 as follows, i.e., $\forall v, v' \in \mathcal{V}, v \neq v'$,
\begin{equation}
\textcolor{black}{\sum_{i \in \mathcal{M}} {\tilde{x}}_{v'i} - \sum_{i \in \mathcal{M}} {\tilde{x}}_{vi} \leq T_{\emph{max}} \cdot \sigma_{vv'} - \varepsilon \cdot (1-\sigma_{vv'})}
\end{equation}
\begin{equation}
\sum_{i \in \mathcal{M}} \tilde{x}_{vi} + p_{v} - \sum_{i \in \mathcal{M}} \tilde{x}_{v'i}\leq  T_{\emph{max}} (2- \sigma_{vv'}- \sum_{i \in M} \boldsymbol{\psi}_{vv'i})
\end{equation}
\subsubsection{Communication resource constraint reformulation}
Similarly, to ensure the data transmission does not conflict over wired links or wireless subchannels, constraints (20)-(23) should be satisfied, i.e., $\forall e,e' \in \mathcal{E}, e \neq e'$, 
\begin{equation}
{\tilde{y}}_{e'b} -  {\tilde{y}}_{eb} \leq T_{\emph{max}} \cdot \sigma_{ee'} -\varepsilon \cdot (1-\sigma_{ee'})
\end{equation}
\begin{equation}
{\tilde{y}}_{eb} + q_{e} -{\tilde{y}}_{e'b}\leq T_{\emph{max}}\cdot  (2-\phi_{ee'} - \boldsymbol{\chi}_{ee'b})
\end{equation}
\begin{equation}
\sum_{k \in {\mathcal{K}}} \tilde{y}_{e'k} - \sum_{k \in {\mathcal{K}}} \tilde{y}_{eb} \leq T_{\emph{max}} \cdot \sigma_{ee'} - \varepsilon \cdot (1-\sigma_{ee'})
\end{equation}
\begin{equation}
\sum_{k \in {\mathcal{K}}} \tilde{y}_{ek} + \textcolor{black}{\check{q}_{e}} - \sum_{k \in {\mathcal{K}}} \tilde{y}_{e'k} \leq  T_{\emph{max}} (2-\phi_{ee'} - \sum_{k \in {\mathcal{K}}}  \boldsymbol{\chi}_{ee'k})
\end{equation}
\subsubsection{Precedence constraints reformulation}
To coordinate the computing task execution and bandwidth augmentation and maintain the consistency of task and data transfer decisions, each task or data transfer can only start after all of its precedent tasks are completed, i.e.,
\begin{equation}
\sum_{i \in \mathcal{M}} {\tilde{x}}_{vi}  + p_{v}  \leq \sum_{k \in \mathcal{K}\cup \{b,c\}} {\tilde{y}}_{(uv),k}
\end{equation}
\begin{equation}
\begin{split}
& \sum_{k \in \mathcal{K}\cup \{b,c\}} {\tilde{y}}_{(uv),k}  + q_{{uv}} {{y}}_{(uv),b}  
+ \check{q}_{{uv}} \sum_{k \in {\mathcal{K}}} {y}_{(uv),k}
\\&  +  r_{uv} {y}_{(uv),c} + \sum_{i \in \mathcal{M}}  {\tilde{x}}_{vi} \leq \sum_{i \in \mathcal{M}} {\tilde{x}}_{vi},
\end{split}
\end{equation}
where $  {{y}}_{(uv),b}  + \sum_{k \in {\mathcal{K}}} {y}_{(uv),k} + {y}_{(uv),c} =1 $ is explicitly guaranteed earlier in constraint (11).   
Additionally, since if the adjacent tasks of an edge $(u,v)$ are assigned to the same rack, the intermediate data will be transferred locally without occupying network resources. Thus, the coupling constraints between the assignment of tasks and data transfer can be written as:
\begin{equation}
\sum_{i \in M} \boldsymbol{\psi}_{uvi} = {y}_{(uv),c}, \forall (u,v) \in \mathcal{E}.
\end{equation}

As such, all of the constraints in \textbf{OP} are linearized and the problem can be reconstructed as follow,
\begin{align*}
\mathbf{RP: }\text{ min }   &  C_{\emph{max}}\\
\text{s.t. }  & (11)-(26),
\\ & {T}_\emph{max} \geq {C}_\emph{max} \geq {T}_\emph{min} \geq \sum_{i \in \mathcal{M}}  \check{x}_{vi} + p_v, \forall v \in {\mathcal{V}}.
\end{align*}

As a result, we transform the MINLP into a linearized one with the help of its bounds and the generalized data transfer model, thus the \textbf{OP} can be solved by solving \textbf{RP}.
Note that, \textbf{OP} and \textbf{RP} are equivalent since the satisfaction of all constraints in \textbf{RP} indicate the satisfaction of the ones of  \textbf{OP}, and vice versa. \textbf{RP} can be optimally solved by the Branch and Bound (\textbf{B\&B}) algorithm \cite{wolsey1999integer}, making it possible to jointly schedule jobs and wireless transceivers efficiently.

\subsection{Decomposition and Acceleration}

To further speed up the solving procedure of using the \textbf{B\&B}, we decompose the \textbf{RP} into multiple feasibility sub-problems. The feasibility sub-problems are derived from \textbf{RP} conditioned on the moving upper bound $T_\emph{max}$, which is formulated as 
\begin{equation*}
\begin{split}
\mathbf{FP: }\text{find }   &  \mathbf{x},\mathbf{\tilde{x}},\mathbf{y},\mathbf{\tilde{y}}
\\ 	\text{s.t. }  & \text{constraints in  }\textbf{RP},
\\ & \text{set }  T_\emph{max} = \ell, \text{and } \ell \in [T_\emph{min},T_\emph{max}],
\end{split}
\end{equation*}
where $\ell$ is the updated upper bound of $C_\emph{max}$.
During each iteration, we assume that the sub-problem is feasible, and start with an interval $[T_\emph{min},T_\emph{max}]$ which is known to contain the optimal solution value $C^*_\emph{max}$. We then solve the feasibility sub-problem at its midpoint $\ell=\frac{T_\emph{min}+T_\emph{max}}{2}$ to determine whether the optimal solution is in the lower or upper half of the interval, and narrow the interval accordingly. Each iteration the interval is bisected, so the width of the interval after $g$ iterations is $2^{-g}(T_\emph{max}-T_\emph{min})$. Repeat this procedure until the width of the interval is small enough. Eventually, the $\mathbf{s}$ in \textbf{OP} can be acquired as $s_v=\sum_{i \in M} \tilde{x}_{vi}, \forall v \in \mathcal{V}$, and $s_{(u,v)}=\sum_{k \in \mathcal{K} \cup \{b,c\}} \tilde{y}_{ek}, \forall (u, v) \in \mathcal{E}$.




\section{Simulation Results}

We implemented the proposed method using Gurobi \cite{gurobi} and evaluated the performance gain introduced by wireless links 
through numerical simulations.
Similar to \cite{giroire2019network}, we randomly generated three types of jobs, i.e., simple MapReduce workflows, one-stage MapReduce workflows and random workflows with computing tasks whose processing time are uniformly chosen from [1,100]. The \textit{network factor} $\rho$, which represents the ratio between the average data transfer time and the average processing time, is defined to set data transfer time. \textcolor{black}{The larger the network factor, the higher the data size.}
As in \cite{halperin2011augmenting, han2015rush, li2019energy}, we assume both wired and wireless links have a transmission rate of 10 Gbps, and focus on scenarios where each of allocated wireless subchannels can fulfill the job's specified bandwidth requirement as the wired links.

In Fig. 4, we compare our method with six different wired-links-only job scheduling baselines in terms of job completion time.
Specifically, the Random Scheduling scheme distributes computing tasks randomly, while the List Scheduling scheme is from \cite{rayward1987uet}. The Partition Scheduling, Generalized List (G-List) Scheduling and G-List-Master Scheduling schemes are from \cite{giroire2019network}. The Optimal Scheduling scheme with only wired links is derived from our method by dropping wireless resources.
We fix the network factor $\rho = 0.5 $ to mimic the scenario where approximately half of the time is spent on data transfers as reported in \cite{chowdhury2011managing}. The task number of each job is chosen from $[5,10]$, aligning with the production job statistics from \cite{ren2012workload} that the majority of jobs contain tasks less than $10$.
It can be observed that when the racks (computing resources) are insufficient, the performance gain introduced by wireless links is relatively small. As the available rack number increases, adding wireless sub-channels 
can reduce the job completion time by up to $10\%$.
However, adding more than one wireless subchannel contributes relatively less to job performance.
\begin{figure}[t!]
	\centering
		\includegraphics[width=85mm]{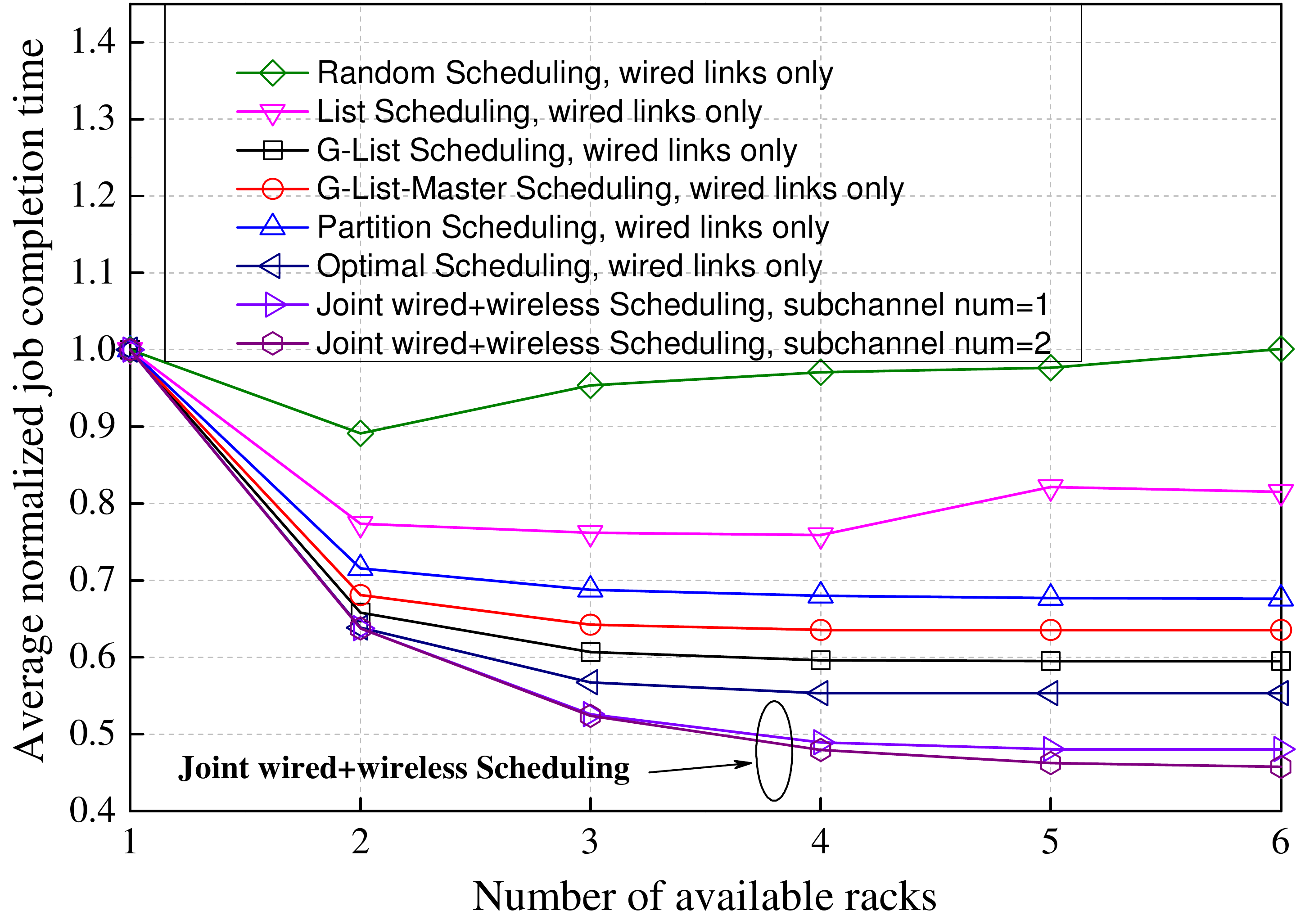}\\	
	\caption{Average job completion time with and without wireless subchannels as a function of the number of available racks for jobs with ten tasks.}
\end{figure}
\begin{figure}[t!]
	\centering
	\includegraphics[width=85mm]{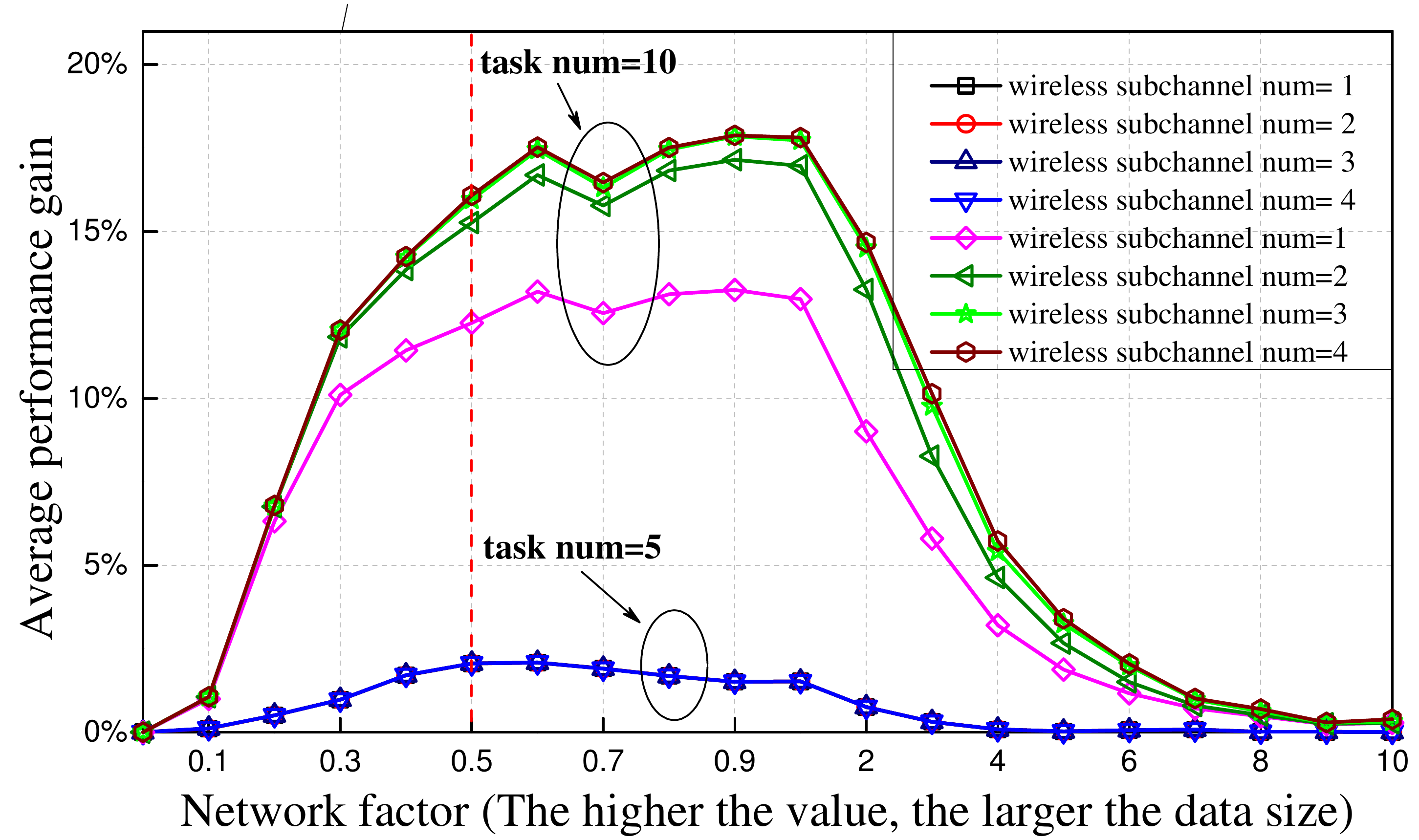}\\
	\caption{Average performance gain of adding wireless resources versus network factor ratio on jobs with different number of tasks.}
\end{figure}

In Fig. 5,  we fix the available rack number as $|\mathcal{V}|$, and vary the network factor from 0.1-10 to show the impact of the increased data size on average performance gain. 
As is seen from the figure, with the increase of the network factor, the performance gain increases at first and then decreases. The reason is when data size is small, the benefit of optimizing data transfer is slight. 
Under this scenario, increased data size may cause worser tardiness and thus wireless augmentation can bring higher benefits. 
As the network factor continues to increase, the data transfer time becomes even longer than task processing time, in this scenario, 
it might be better to assign all computing tasks of a job to a single rack to avoid data transfer.
Besides, with fixed network factor (e.g., the red dashed vertical line), the larger the task number, the higher the performance gain can be achieved by wireless bandwidth augmentation. 
And adding more wireless resources brings reduced gains.

\section{Conclusion}
In this work, we investigated the joint job scheduling and bandwidth augmentation in hybrid data centers. We observed the wireless-augmented job scheduling problem is an MINLP, which is not solvable by existing optimization methods. Thus, we linearized the original model with help of its bounds, the revised data transfer model and the disjunctive reformulation technique, such that it can be solved optimally by the Branch and Bound method. Simulation results showed that jointly scheduling the tasks and wireless transceivers can significantly reduce the job completion time. In our future work, we will study job scheduling problems that involving more real-world constraints for online scenarios.
\bibliographystyle{IEEEtran}
\bibliography{reference}

\begin{thebibliography}{10}
\providecommand{\url}[1]{#1}
\csname url@samestyle\endcsname
\providecommand{\newblock}{\relax}
\providecommand{\bibinfo}[2]{#2}
\providecommand{\BIBentrySTDinterwordspacing}{\spaceskip=0pt\relax}
\providecommand{\BIBentryALTinterwordstretchfactor}{4}
\providecommand{\BIBentryALTinterwordspacing}{\spaceskip=\fontdimen2\font plus
\BIBentryALTinterwordstretchfactor\fontdimen3\font minus
  \fontdimen4\font\relax}
\providecommand{\BIBforeignlanguage}[2]{{%
\expandafter\ifx\csname l@#1\endcsname\relax
\typeout{** WARNING: IEEEtran.bst: No hyphenation pattern has been}%
\typeout{** loaded for the language `#1'. Using the pattern for}%
\typeout{** the default language instead.}%
\else
\language=\csname l@#1\endcsname
\fi
#2}}
\providecommand{\BIBdecl}{\relax}
\BIBdecl

\bibitem{dean2008mapreduce}
J.~Dean and S.~Ghemawat, ``Map{R}educe: Simplified data processing on large
  clusters,'' \emph{Commun. ACM}, vol.~51, no.~1, pp. 107--113, 2008.

\bibitem{malewicz2010pregel}
G.~Malewicz, M.~H. Austern, A.~J. Bik, J.~C. Dehnert, I.~Horn, N.~Leiser, and
  G.~Czajkowski, ``Pregel: {A} system for large-scale graph processing,'' in
  \emph{Proc. ACM SIGMOD Int. Conf. on Manage. Data}, 2010, pp. 135--146.

\bibitem{zaharia2012resilient}
M.~Zaharia, M.~Chowdhury, T.~Das, A.~Dave, J.~Ma, M.~McCauly, M.~J. Franklin,
  S.~Shenker, and I.~Stoica, ``Resilient distributed datasets: A
  {Fault-Tolerant} abstraction for {In-Memory} cluster computing,'' in
  \emph{Proc. USENIX Symp. Netw. Syst. Design Implement. (NSDI)}, 2012, pp.
  15--28.

\bibitem{chowdhury2011managing}
M.~Chowdhury, M.~Zaharia, J.~Ma, M.~I. Jordan, and I.~Stoica, ``Managing data
  transfers in computer clusters with orchestra,'' in \emph{Proc. ACM SIGCOMM
  Conf.}, vol.~41, no.~4, 2011, pp. 98--109.

\bibitem{terzi202160}
C.~Terzi and I.~Korpeoglu, ``60 {GHz} wireless data center networks: A
  survey,'' \emph{Computer Networks}, vol. 185, p. 107730, 2021.

\bibitem{celik2019optical}
A.~Celik, B.~Shihada, and M.-S. Alouini, ``Optical wireless data center
  networks: potentials, limitations, and prospects,'' in \emph{Broadband Access
  Commun. Tech. XIII}, vol. 10945.\hskip 1em plus 0.5em minus 0.4em\relax
  International Society for Optics and Photonics, 2019, p. 109450I.

\bibitem{abari2016millimeter}
O.~Abari, H.~Hassanieh, M.~Rodriguez, and D.~Katabi, ``Millimeter wave
  communications: From point-to-point links to agile network connections,'' in
  \emph{Proc. ACM Workshop Hot Topics Netw.}, 2016, pp. 169--175.

\bibitem{ghobadi2016projector}
M.~Ghobadi, R.~Mahajan, A.~Phanishayee, N.~Devanur, J.~Kulkarni, G.~Ranade,
  P.-A. Blanche, H.~Rastegarfar, M.~Glick, and D.~Kilper, ``Projector: Agile
  reconfigurable data center interconnect,'' in \emph{Proc. ACM SIGCOMM Conf.},
  2016, pp. 216--229.

\bibitem{luo2019energy}
M.~Luo, J.~Li, J.~Ma, H.~Li, and M.~Sheng, ``Energy-efficient flow routing and
  scheduling in hybrid data center networks,'' in \emph{Proc. IEEE Global
  Commun. Conf. (GLOBECOM)}, 2019, pp. 1--6.

\bibitem{han2015rush}
K.~Han, Z.~Hu, J.~Luo, and L.~Xiang, ``Rush: Routing and scheduling for hybrid
  data center networks,'' in \emph{Proc. IEEE Conf. on Comput. Commun.
  (INFOCOM)}.\hskip 1em plus 0.5em minus 0.4em\relax IEEE, 2015, pp. 415--423.

\bibitem{halperin2011augmenting}
D.~Halperin, S.~Kandula, J.~Padhye, P.~Bahl, and D.~Wetherall, ``Augmenting
  data center networks with multi-gigabit wireless links,'' \emph{ACM SIGCOMM
  Comput. Commun. Rev.}, pp. 38--49, 2011.

\bibitem{cui2013dynamic}
Y.~Cui, H.~Wang, X.~Cheng, D.~Li, and A.~Yl{\"a}-J{\"a}{\"a}ski, ``Dynamic
  scheduling for wireless data center networks,'' \emph{IEEE Trans. Parallel
  Distrib. Syst.}, vol.~24, no.~12, pp. 2365--2374, 2013.

\bibitem{li2019energy}
T.~Li and S.~Santini, ``Energy-aware coflow and antenna scheduling for hybrid
  server-centric data center networks,'' \emph{IEEE Trans. Green Commun.
  Netw.}, vol.~3, no.~2, pp. 356--365, 2019.

\bibitem{ao2021joint}
W.~C. Ao, P.-H. Huang, and K.~Psounis, ``Joint workload distribution and
  capacity augmentation in hybrid datacenter networks,'' \emph{IEEE/ACM Trans.
  Netw.}, vol.~29, no.~01, pp. 120--133, 2021.

\bibitem{ren2012workload}
Z.~Ren, X.~Xu, J.~Wan, W.~Shi, and M.~Zhou, ``Workload characterization on a
  production hadoop cluster: A case study on taobao,'' in \emph{Proc. IEEE Int.
  Symp. Workload Characterization (IISWC)}, 2012, pp. 3--13.

\bibitem{zhang2014fuxi}
Z.~Zhang, C.~Li, Y.~Tao, R.~Yang, H.~Tang, and J.~Xu, ``Fuxi: a fault-tolerant
  resource management and job scheduling system at internet scale,'' in
  \emph{Proc. VLDB Endowment}, vol.~7, no.~13.\hskip 1em plus 0.5em minus
  0.4em\relax VLDB Endowment Inc., 2014, pp. 1393--1404.

\bibitem{wolsey1999integer}
L.~A. Wolsey and G.~L. Nemhauser, \emph{Integer and combinatorial
  optimization}.\hskip 1em plus 0.5em minus 0.4em\relax John Wiley \& Sons,
  1999, vol.~55.

\bibitem{gurobi}
\BIBentryALTinterwordspacing
L.~Gurobi~Optimization, ``Gurobi optimizer reference manual,'' 2021. [Online].
  Available: \url{http://www.gurobi.com.}
\BIBentrySTDinterwordspacing

\bibitem{giroire2019network}
F.~Giroire, N.~Huin, A.~Tomassilli, and S.~P{\'e}rennes, ``When network
  matters: Data center scheduling with network tasks,'' in \emph{Proc. IEEE
  Conf. on Comput. Commun. (INFOCOM)}, 2019, pp. 2278--2286.

\bibitem{rayward1987uet}
V.~J. Rayward-Smith, ``Uet scheduling with unit interprocessor communication
  delays,'' \emph{Discrete Applied Mathematics}, vol.~18, no.~1, pp. 55--71,
  1987.

\end{thebibliography}

\end{document}